\documentclass[showpacs,prb,superscriptaddress,twocolumn]{revtex4-1}
\usepackage[table]{xcolor}
\usepackage{textcomp}
\usepackage{amsmath}
\usepackage{amssymb}
\usepackage{}
\pdfoutput=1
\usepackage{graphicx} 
\usepackage{}
\usepackage{pgf}
\usepackage{hyperref}
\usepackage{epstopdf}
\usepackage{braket}

\newcommand{\e}{\varepsilon}

\begin{document}

\title{Crossover from fast to slow dynamics in quantum Ising chains with long range interactions.}

\author{Giulia Piccitto}
\affiliation{SISSA, Via Bonomea 265, I-34135 Trieste, Italy}

\author{Alessandro Silva}
\affiliation{SISSA, Via Bonomea 265, I-34135 Trieste, Italy}

\begin{abstract}
	Quantum many body systems with long range interactions are known to display many fascinating phenomena experimentally observable in trapped ions, Rydberg atoms and  polar molecules. Among these are dynamical phase transitions which occur after an abrupt quench in spin chains with interactions decaying as $r^{-\alpha}$ and whose critical dynamics depend crucially on the power $\alpha$: for systems with  $\alpha<1$ the transition is sharp while for $\alpha>1$ it fans out in a chaotic crossover region. In this paper we explore the fate of critical dynamics in Ising chains with long-range interactions when the transverse field is ramped up with a finite speed. While for abrupt quenches we observe a chaotic region that widens as $\alpha$ is increased, the width of the crossover region diminishes as the time of the ramp increases, suggesting that chaos will disappear altogether and be replaced by a  sharp transition in the adiabatic limit.   
\end{abstract}

\maketitle
\section{Introduction}\label{sec:introduzione}
In the last decades the engineering of long-range interactions in low dimensional quantum many-body systems has been the focus of many experiments, ranging from Rydberg atoms \cite{Rydberg:1, Rydberg:2, Rydberg:3} to polar molecules \cite{Polar:1, Polar:2} and trapped ions \cite{Trapped:1,Trapped:2,Trapped:3,Trapped:4}. These systems give the possibility to study a rich variety of phenomena which contrast the non-equilibrium dynamics of systems with long-range interactions to that of short-range ones, such as the violation of Lieb-Robinson bounds and anomalous propagation of information~\cite{nonlocal:exp:1,entanglement:exp:1,entanglement:the:1, entanglement:the:2,Lieb:1, Lieb:2, Lieb:3}, the localization of kink-like excitations~\cite{confined:1,confined:2} as well as the observation of prethermal phases and dynamical phase transitions~\cite{neyenus,dqpt:1,dqpt:2, dqpt:3, dqpt:4, dqpt:5, dqpt:6,halimeh:1}. 

It is known that dynamical phase transitions are very sensitive to the range of interactions~\cite{dqpt:3}, in fact moving from short range to long range ones new dynamical features emerge. In a recent work \cite{giulia:1} we have shown that for power law decaying interactions ($J(r) \propto 1/r^{\alpha}$) the dynamical properties are very sensitive to the precise value of the power law exponent $\alpha$. In particular, for quantum Ising chains subject to a quench in the transverse field the dynamical phase transition for  $\alpha<1$ is sharp, while for $\alpha>1$ the critical point fans out in a crossover chaotic region where the dynamics and the asymptotic state depend sensitively on the system parameters ~\cite{giulia:1}.
While the overall dynamics is dominated by long-range correlations among spins, the key ingredient to observe chaos appear to be residual short range correlations~\cite{giulia:1,dqpt:7,dqpt:8} which effectively damp the order parameter dynamics. In the chaotic region, the spin system behaves collectively like a tossed coin:  initially the magnetization flips periodically  between positive and negative values along paramagnetic trajectories, until the damping produces relaxation  and localization on stable ferromagnetic trajectories with pseudo-random asymptotic magnetization~\cite{giulia:1, dqpt:7, dqpt:8}. 

The robustness of this chaotic region can be investigated by taking out of equilibrium the system with a linear ramp. Changing the control parameter with a finite velocity, in fact, can affect the interplay between nonequilibrium dynamics and dissipation suppressing chaos in favour of regular dynamics. Intuitively, while an abrupt quench is analogous to a fast coin toss, varying smoothly the system parameters is like placing slowly a coin on the table, a process that is obviously devoid of uncertainty.  In this work we will show that decreasing the ramping speed of the transverse field in an Ising model with long range interactions has the same effect: the chaotic region shrinks in size until the dynamical phase diagram resembles the equilibrium one in the adiabatic limit.   
We will show this using the cluster mean field theory (CMFT) \cite{cmf:1, cmf:2}. In the following we first present the model, a quantum Ising chain with power law decaying interactions, and its equilibrium phase diagram (Sec. \ref{sec:equilibrium}). Then we investigate the dynamics when the transverse field is smoothly switched on with a particular focus on the robustness of the chaotic region and on the crossover between randomness and predictability (Sec. \ref{sec:rampa}).

\section{Model and equilibrium phase diagram}\label{sec:equilibrium}
Let us first introduce the model. In this paper we study the long range interacting quantum Ising chain, described by the following Hamiltonian
\begin{equation}
	H = -\frac{J}{N(\alpha)}\sum_{i\ne j}^N  \frac{\sigma^z_i \sigma^z_j}{|i-j|^\alpha} - h \sum_i^N \sigma^x_i,
	\label{model}
\end{equation}
where $\sigma^\beta_i$ are Pauli matrices acting on site $i$,  $N(\alpha) = \sum_{r=1}^N r^{-\alpha}$ the usual Kac normalization.
In Ref. [\onlinecite{eq:ddf:1}] it was shown that in the 1D case the quantum phase transition is present for all transverse fields but the critical exponent at zero temperature are described by mean field theory only for $\alpha \le \alpha_c = 5/3$. In turn, a pure mean field description would predict, for all $\alpha$, a sharp equilibrium phase transition from a ferromagnetic to a paramagnetic phase at $h = 2J$. Mean field theory is in turn exact for $\alpha <1$, therefore it can be used as a benchmark on the validity of any other approximation.
In order to get a feeling of what happens for larger $\alpha$ we decided to use the cluster mean field introduced in Ref. [\onlinecite{cmf:1}, \onlinecite{ cmf:2}] that accounts for the effects of short range correlations.  
The idea is to divide the system into $N_\text{cl}$ clusters of size $\ell$ living in the mean field generated by others and to deduce the physics of the system from the exact physics of one of these clusters. 
As shown in Ref. [\onlinecite{giulia:1}] the CMFT allows one to write the Hamiltonian as the sum of two parts $H = H_{\text{cl}} + H_{\text{mf}}$, with
\begin{equation}
\begin{aligned}
	&H_{\text{cl}} = -\frac{1}{N(\alpha)}\sum_\beta^{N_{\text{cl}}}\left( \sum_{i < j \in \beta}^L  \frac{\sigma_i^z 	\sigma_j^z }{|i-j|^\alpha}- h \sum_{i \in \beta}^L \sigma^x_i\right)\\
	&H_{\text{mf}} = 2 \bar{m} \sigma^z J_{\text{eff}},
	\label{ham_mf}
\end{aligned}
\end{equation}
where $\beta$ is the cluster index, $\bar{m} = \frac{1}{L}\sum_{i \in \beta}^L \braket{\sigma^z_i}$ the mean value of the magnetization inside the cluster $\beta$, $J_{\text{eff}} = \frac{1}{N(\alpha)} \sum_{n = 1}^{N_{cl}} \frac{J}{|nL|^\alpha}$ and $\sigma^z = \sum_i \sigma^z_i$. 
In the whole article we will consider the ferromagnetic case $J = 1$ only and we will set $\hbar = 1$. 

To derive the equilibrium phase diagram we solve self consistently for the ground state of the Hamiltonian \eqref{ham_mf} with a precision $\e = 1e-5$. 
\begin{figure}[h!]
	\centering
	\includegraphics[scale = 0.37]{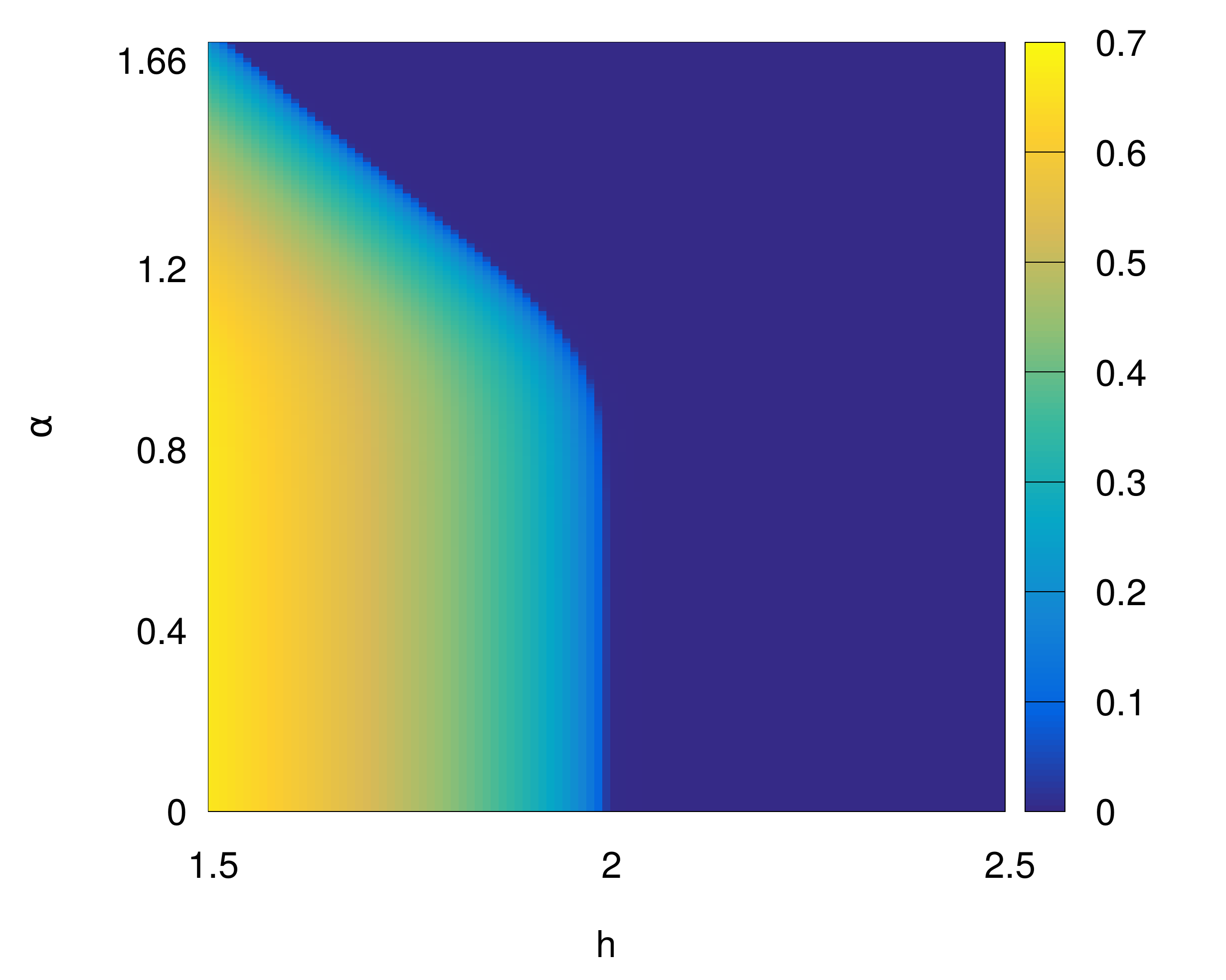}
	\caption{(Color online). Equilibrium phase diagram for the quantum long range Ising model in the cluster mean field approximation obtained with a cluster size $\ell = 5$. The order parameter $ \frac{1}{\ell}\sum_i^\ell\braket{ \sigma^z}$ (color scale) as a function of the transverse field $h$ and the power law exponent $\alpha$. A quantum phase transition from a ferromagnetic to a paramagnetic phase occurs at a critical value $h_c(\alpha)$. What emerges is that for $\alpha < 1 $ the critical field coincide with the mean field one while in the other case it assumes a dependence on the power law exponent.}
	\label{ddf_eq} 
\end{figure}
We chose the cluster length $\ell = 5$.
The result is shown in Fig. \ref{ddf_eq} where the order parameter $m = \frac{1}{\ell}\sum_i^\ell\braket{ \sigma^z}$ is plotted as a function of the transverse field $h$ and the power law exponent $\alpha$. 
This result is in perfect agreement with the exact solution of the model for $\alpha < 1$ \cite{lmg:1}.
In the regime $\alpha >1$ we observe that the quantum critical point $h_c(\alpha)$ assumes a linear dependence on the power law exponent. 
From this observation it is possible to extract a qualitative behavior of the quantum critical point as a function of $\alpha$. 
Interpolating the data we find
\begin{equation}
	h_c(\alpha ) \sim \begin{cases} 2 \qquad \qquad \qquad\   \text{ if } \alpha \le1, \\ 0.7(3.2 - \alpha ) \quad \  \ \text{ if } 1 < \alpha < 2. \end{cases}
\end{equation}
Since our method is essentially a generalized mean field theory we will use it when the latter describes the equilibrium transition ($\alpha < 5/3$). Despite this, it must be stressed that CMFT is exact for large $\ell$.

\section{Linear quench} \label{sec:rampa}

Now that we have the equilibrium phase diagram to benchmark our non-equilibrium investigations, let us focus on the dynamics following a linear quench of the transverse field. We have already shown in Ref. \onlinecite{giulia:1} that the asymptotic state attained after a sudden quench strongly depends on the value of the power law exponent $\alpha$. For $\alpha < 1$ the system exhibits the mean-field dynamical quantum phase transition from a dynamical ferromagnetic to a dynamical paramagnetic phase at $h = J$, as predicted in the literature \cite{lmg:1}. As soon as $\alpha >1$ this critical point spreads in a critical region where the system exhibits chaotic behavior and a strong dependence on the initial conditions.
To investigate the robustness of this phenomenon we study the post linear quench dynamics, considering the Hamiltonian with a transverse field varied according to
\begin{equation}
	h(t) = \begin{cases} 0 \qquad \qquad \quad \text{ if } t < 0, \\ h\tanh (\lambda t) \quad \text{ if } t \ge 0. \end{cases} 
\end{equation}
The limit $\lambda \to \infty$ coincides with the sudden quench dynamics we have already described. 
Jaschke and collaborators \cite{kz:1} have shown that the physics of the Kibble-Zurek mechanism holds despite the long range interactions, thus we expect to find in the adiabatic limit $\lambda \to 0 $ the same phase diagram as in in Fig. \ref{ddf_eq}. 
For intermediate values of the slope we expect to obtain informations on the crossover between the chaotic and regular dynamics. 

\begin{figure*}[t!]
	\centering
	\includegraphics[scale = 0.31]{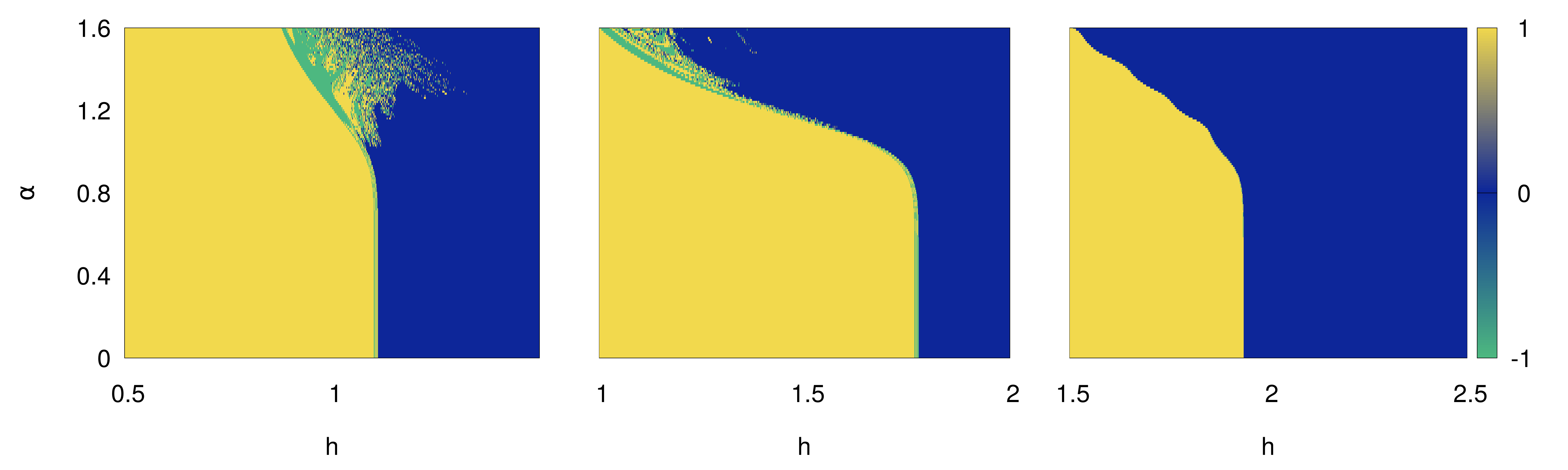}
	\caption{(Color online). Sign of long-time averaged longitudinal magnetization $m^T$ (color scale) as a function of the power law exponent $\alpha$ and the final transverse field $h$ for the different slopes $\lambda = 1, 0.5, 0.05$ (respectively from left to right).
	Yellow points indicate positive magnetization, green points indicate negative magnetization, blue points indicate zero magnetization. Three different regimes can be observed. In the left panel we observe a dynamics that is very close to the post (sudden) quench one and the chaotic behavior is still present. In the right panel we observe the limit $\lambda \to 0$ and the dynamics becomes adiabatic, in fact the equilibrium phase diagram is recovered. Finally, in the central panel we can observe an intermediate regime in which the dynamics is going towards the equilibrium but some chaotic reminiscence is still present.}
	\label{ddf} 
\end{figure*}
\begin{figure}
	\includegraphics[scale = 0.3]{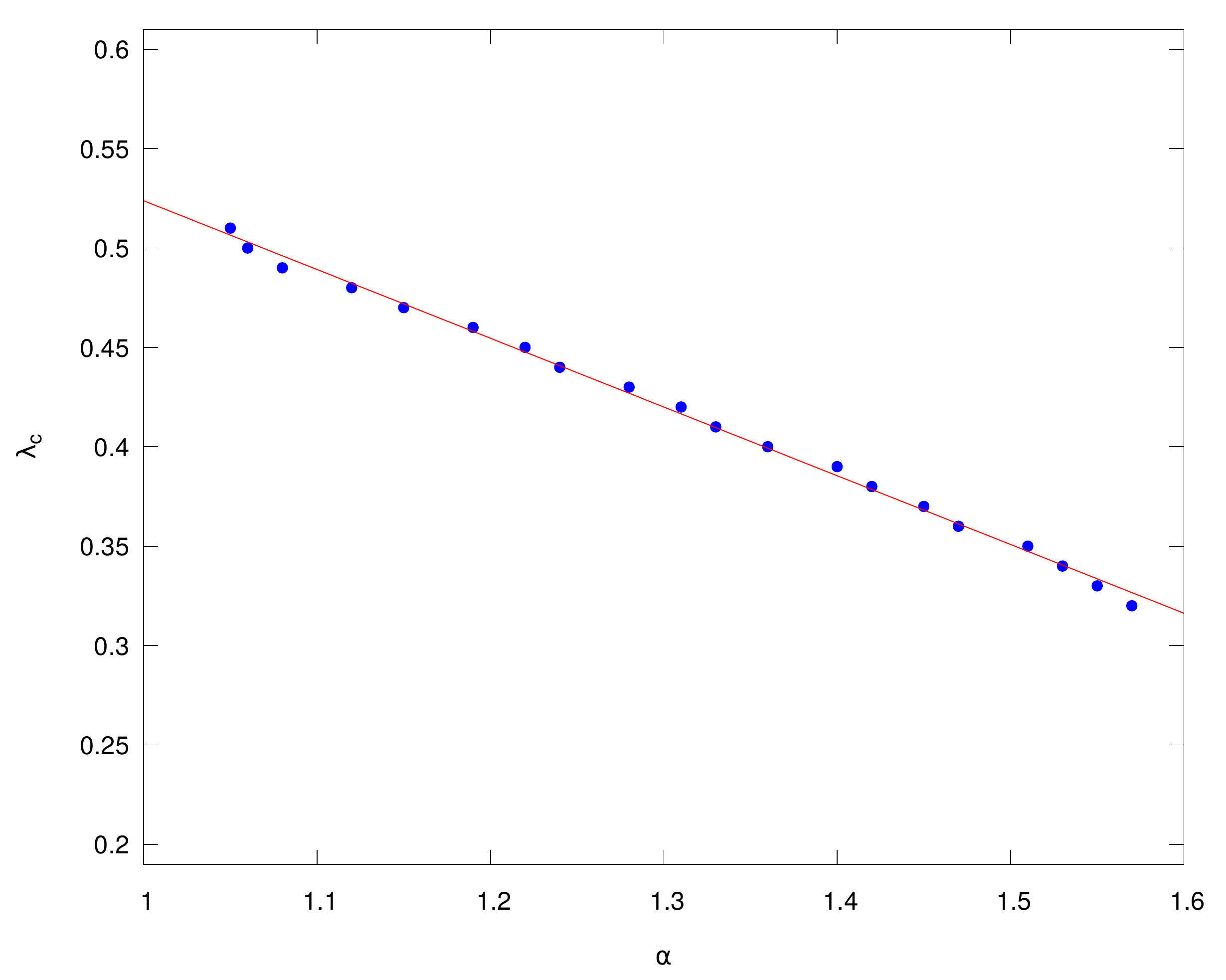}
	\caption{In this figure we show the behavior of $\lambda_c$ as a function of $\alpha$. The blue point are the data extrapolated from the numerical results. The red line is the best fit interpolating them. From this figure we can observe a linear trend confirming the claim that the higher the power law exponent the slower the ramp should be to obtain a sharp phase transition.}
	\label{lc}
\end{figure}

To this purpose, we simulated the linear quench dynamics using the cluster mean field approach.
The system is initially prepared in the ground state of the Hamiltonian $H_0 = \sum_{ij} J_{ij} \sigma^z_i \sigma^z_j$, i.e. all spins polarized along the $z$ direction, and at time $t = 0$ the transverse field $h(t) = h \tanh(\lambda t)$ is turned on.
The dynamics has been derived integrating self consistently the Schr\"odinger equation using an explicit embedded Runge-Kutta-Fehlberg(4,5) step adaptive method\footnote{We used the routine of runge-kutta integration in the GNU scientific library that can be found at https://www.gnu.org/software/gsl/doc/html/ode-initval.html}. We have fixed for all the simulations $N_{\text{cl}} = 10^7$. The self consistent field $\overline{m}$ is update every $dt = 1e-3$.   
The order parameter we are interested in is the long-time averaged longitudinal magnetization $m^T = \lim_{T \to \infty} \frac{1}{T} \int_{t^*}^T dt \braket{\sigma^z(t)}$. 
The main result can be summarized in the phase diagram in Fig. \ref{ddf} that shows the sign of $m^T$ as a function of the power law exponent $\alpha$ and the transverse field $h$ for three different value of $\lambda = 1, 0.5, 0.05$. The simulation has been run with $\ell = 5$ evolving the dynamics up to a time $T = 200$. The order parameter has been evaluated averaging in the time window $t \in [180, 200]$ after the asymptotic state has been reached. 
The times are expressed in unity of $J$.
From this result we deduce that there are three different regimes. 
The first (left panel) is the one of the sharp ramp in which, except for a small shift of the dynamical critical point, we recover the same dynamical phase diagram  as in Ref. \onlinecite{giulia:1} with the same chaotic features. 
The second one (right panel) is the limit $\lambda \to 0$ in which, as expected, the chaos is absent and the phase diagram starts to resemble the equilibrium one with a slight shift in the critical point.
The last regime can be observed for intermediate value of the ramp slope. As can be observed in the central panel of Fig. \ref{ddf}, in this regime the system is slowly moving towards the equilibrium phase diagram but still displays a chaotic phase. 
From this analysis is still not clear how the crossover between the sudden quench and the adiabatic regime occurs. However, we can analyze the fine details of the phase diagrams close to the critical point to extract some qualitative information on the behavior of the chaotic region. In analogy with what done in Ref. \onlinecite{coin:1, giulia:1}, we define the maximum size of the neighborhood $\varepsilon$ of points in the parameter space of the same phase, chaos is defined by the condition $\varepsilon  = 0$. In particular, for a fixed value $\alpha$, $\varepsilon(h)$ is the size of the biggest square centered in h containing points with the same sign of the order parameter. 
\begin{figure*}
	\centering
	\includegraphics[width = \textwidth]{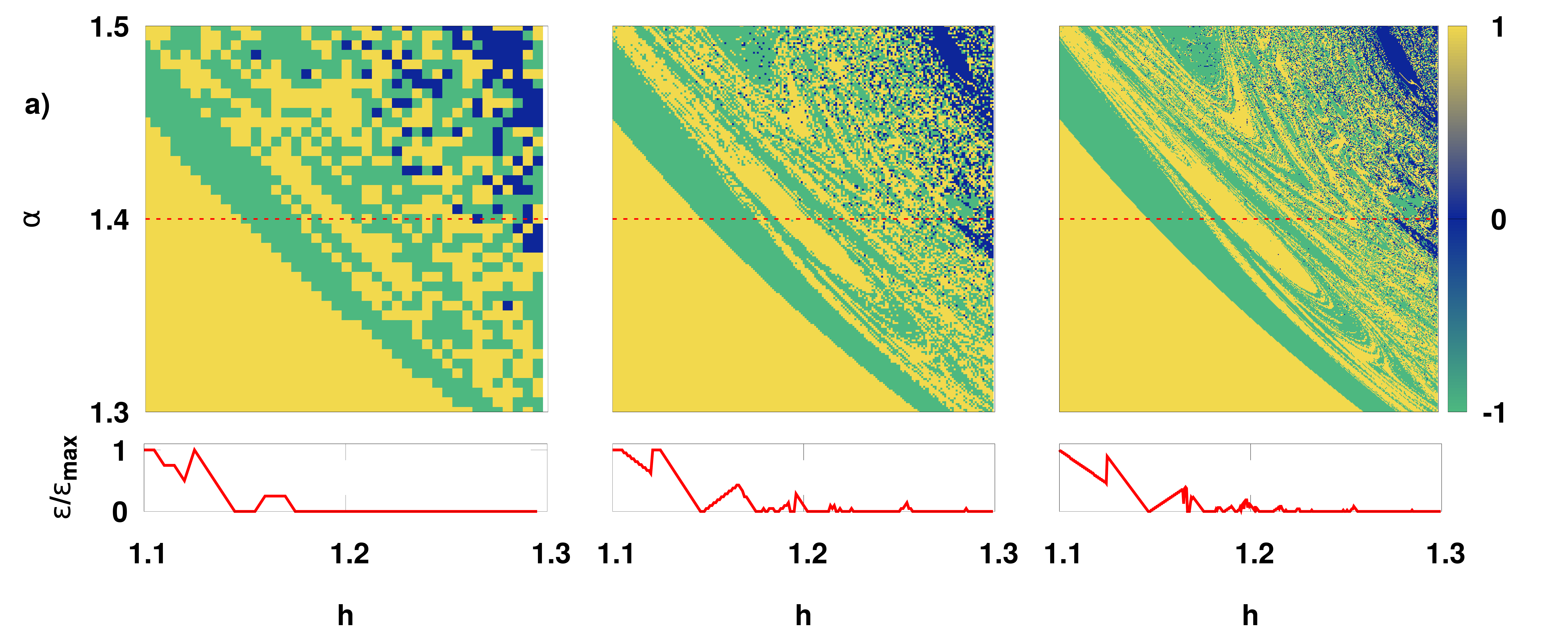}
	\includegraphics[width = \textwidth]{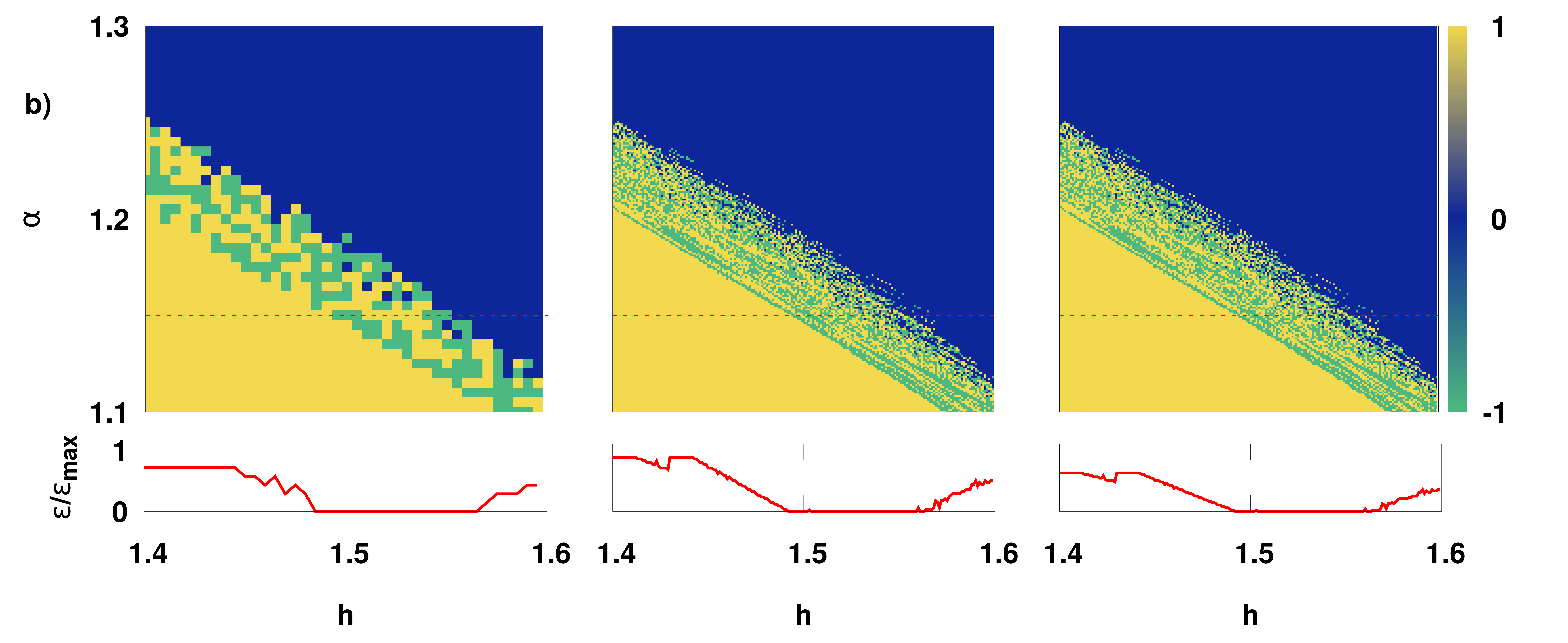}
	\caption{(color online). \textbf{a)} Upper panels: a portion of the phase diagram ($1.3 < \alpha < 1.5$ and $1.1 < h < 1.3$) for three different resolutions (from the left to the right: $\delta \alpha = \delta h = 0.005, 0.001, 0.0005$). Bottom panels: normalized neighborhood $\varepsilon(h)/\max_h \varepsilon(h)$ evaluated at $\alpha = 1.4$ (red dotted line). It emerges that for these values of the power law exponent $\varepsilon \to 0$ and the system preserve the chaotic features displayed in the case of the sudden quantum quench.
	\textbf{b)} Upper panels: a portion of the phase diagram ($1.1 < \alpha < 1.3$ and $1.4 < h < 1.6$) for three different resolutions (from the left to the right: $\delta \alpha = \delta h = 0.005, 0.001, 0.0005$). Bottom panels: normalized neighborhood $\varepsilon(h)/\max_h \varepsilon(h)$ evaluated at $\alpha = 1.15$ (red dotted line). It emerges that for this value of the power law exponent the region in which $\varepsilon \to 0$ shrinks, sign of a regularization of the dynamics.}
	\label{caos:1}
\end{figure*}
In the upper panels of Fig. (\ref{caos:1}a) we plot simulations obtained with increasing resolutions (from the left to the right: $\delta \alpha = \delta h = 0.005, 0.001, 0.0005$) of a portion of the phase diagram ($1.3 < \alpha < 1.5$ and $1.1 < h < 1.3$). In the lower panels we plot the respective normalized neighborhood $\varepsilon(h)/\max_h \varepsilon(h)$ evaluated at $\alpha = 1.4$ (red dotted line). It emerges that for these values of the power law exponent $\varepsilon \to 0$ and the system preserve the chaotic features displayed in the case of the sudden quantum quench. 
When we move toward smaller values of $\alpha$ we can see that the chaotic region shrinks. This can be observed in the Fig. (\ref{caos:1}b) where the portion of the phase diagram with $1.1 < \alpha < 1.3$ and $1.4 < h < 1.6$ is plotted as a function of $\alpha$ and $h$. In the bottom panels the quantity $\varepsilon(h)/\max_h \varepsilon(h)$, evaluated along the line $\alpha = 1.15$, is plotted as a function of the final transverse field. It emerges that the region in which $\varepsilon \to 0$ is smaller, sign that chaos is slowly breaking down. 
From this analysis we can qualitatively argue that the bigger the power law exponent the more robust is chaos. Therefore, we can conclude that the crossover between the chaotic and the regular dynamics will start first from small power law exponents and will move toward the bigger ones. 
	A quantitative analysis of this behavior can be obtained by looking at the critical value $\lambda_c$ of the slope, at fixed $\alpha$, below which the transition is sharp. In Fig. \ref{lc} we plot $\lambda_c$ as a function of the power law exponent. What emerges is a linear relation between $\lambda_c$ and $\alpha$. This result confirms the intuition that the bigger $\alpha$ the smoother has to be a quench in order to observe a sharp phase transition.

	Finally we want to stress that we expect the chaotic region disappears for $\alpha > 2$ in agreement with the expectations short range Ising model can not support dynamical phase transitions. Unfortunately, the adopted method does not allow us to investigate the crossover at $\alpha= 2$, thus we can not derive details on the dynamics for shorter interactions.
\section{Conclusion}
In this paper we have presented the crossover from fast to slow dynamics in the quantum Ising chain with long range interactions. 
First we used the cluster mean field theory to derive self consistently the equilibrium phase diagram in the limit $\alpha >1 $ obtaining an hint on how the crossover from long range to short range should occurs. Then we used the same method to simulate the post linear quench dynamics.
The results suggest that there are three different regimes of the dynamics that can be observed. The first one, for sharp ramp, exhibits a dynamics that is very close to the post (sudden) quench one and presents the same chaotic features. In the limit of infinite slow quench we are in the adiabatic limit and we recover the equilibrium phase diagram. The last one is an intermediate regime in which the system is slowly going toward a regular phase but some chaotic islands are still present.
Studying the robustness of the chaotic region as a function of $\alpha $ and $\lambda$ we found that the larger is $\alpha$ the slower has a ramp to be to avoid the chaotic region.

\section{Aknowledgment}
We thank A. Dutta for discussions. G. Piccitto wants to thank L. Arceci for useful discussions.

\bibliographystyle{ieeetr}
\bibliography{rampa}

\end{document}